\documentclass[11pt]{article}
\usepackage{epsfig}
\usepackage[usenames]{color}
\usepackage{graphicx}
\usepackage{amsmath}
\def\la{\langle}\def\ra{\rangle}
\def\be{\begin{eqnarray}}\def\ee{\end{eqnarray}}
\def\lsim{\mathrel{\rlap{\lower3pt\hbox{\hskip1pt$\sim$}}
     \raise1pt\hbox{$<$}}} %less than or approx. symbol
\def\gsim{\mathrel{\rlap{\lower3pt\hbox{\hskip1pt$\sim$}}
     \raise1pt\hbox{$>$}}} %greater than or approx. symbol
\def\le{ \begin{array}{ll}}\def\re{\end{array}}

\def\lear{ \left( \begin{array}{cc}}\def\rear{\end{array} \right)}

\def\le{ \left( \begin{array}{cc}}\def\re{\end{array} \right)}

\def\bi{\bibitem}

\newcommand{\Rmnum}[1]{\expandafter\@slowromancap\romannumeral #1@}
\renewcommand{\thefootnote}{\fnsymbol{footnote}}

\def\lsim{\mathrel{\rlap{\lower3pt\hbox{\hskip1pt$\sim$}}
     \raise1pt\hbox{$<$}}} %less than or approx. symbol
\def\gsim{\mathrel{\rlap{\lower3pt\hbox{\hskip1pt$\sim$}}
     \raise1pt\hbox{$>$}}} %greater than or approx. symbol

\def\la{\langle}
\def\ra{\rangle}
\def\bi{\bibitem}

\begin{document}

\centerline{\Large \bf A New Topological ``Twist" to BR Scaling\footnote{Based on talk given at ``45 Years of Nuclear Theory at Stony Brook--
A Tribute to Gerry Brown", Stony Brook University, November 24-26, 2013}}
\vskip 0.5cm

\centerline{\bf Mannque Rho}

\centerline{\bf\it Institut de Physique Th\'eorique, CEA Saclay, 91191 Gif-sur-Yvette, France}

\centerline{\bf\it \& WCU, Hanyang University, Seoul 133-791, Korea }

\vskip 1cm
\centerline{Abstract}
\vskip 0.3cm
When vector mesons are considered on the same footing as pions as suggested by hidden local symmetry, the property of the nuclear tensor forces is strongly controlled by the behavior of the vector mesons in dense medium. This led to BR scaling in 1991. When baryons as skyrmions are put on crystal, there can be a phase transition from skyrmions to half-skyrmions at a density above that of normal nuclear matter. This topology change can induce fundamental changes to the parameters in hidden local symmetric Lagrangian, hence BR scaling, and brings a drastic modification to the structure of nuclear forces, in particulrar, the tensor forces. This can have far-reaching consequences on the EoS of compact-star matter and the structure of neutron-rich nuclei.
\setcounter{footnote}{0}
\renewcommand{\thefootnote}{\#\arabic{footnote}}
%%%%%%%%%%%%%%%%%%%%%%%%%%%%%%%%%%%%%%%%%%%%%%%%%%%%%

\section{Introduction with a brief history}
In this talk given in a tribute to Gerry Brown, I would like to recount a very recent development in which Gerry was not directly involved but figured eminently in the conceptual underpinning of the problem.

To briefly give a relevant background to the main theme of my talk, let me start when and how my collaboration with Gerry, which lasted nearly 40 years with some 70 papers written, began.  Viewed from my side, our collaboration was triggered by my discussions with Benjamin Lee who was on leave in 1968 from Stony Brook at Institut des Hautes \'Etudes Scientifiques at Bures-sur-Yvette very near Saclay where I was working. He was then working on renormalizing linear sigma model, presaging the breakthrough on the renormalization of  Yang-Mills gauge theory that came later. In my conversations with him, Ben Lee was stressing the possible relevance of chiral dynamics, then in the infant stage with the developments of low-energy theorems (such as soft-pion theorems), in the field theory of strong interaction processes in nuclear physics. I implemented Ben Lee's suggestion by joining in the work Marc Chemtob was doing at Saclay as his thesis, namely, on exchange currents in nuclear weak and electromagnetic currents. The idea was to incorporate the soft-pion theorems in organizing the meson exchange currents that contribute to nuclear electro-weak transition matrix elements.  Gerry Brown, independently working on similar problems focusing mainly on  nuclear forces, was prompted by the paper of Chemtob and me~\cite{chemtob-rho}, and quickly solved with Dan Olof Riska a long-standing problem in nuclear physics, the thermal neutron capture $n+p\rightarrow d+\gamma$, predicting in an extremely simple way the cross-section within a $\sim$ 10\% accuracy~\cite{riska-brown}. What transpired since that time is that the soft-pion terms constituted the leading order term in the exchange current treated in chiral perturbation theory with the next order correction suppressed by two orders of chiral expansion in the magnetic dipole transition involved there~\cite{MR91}. {This observation was dubbed later as ``chiral filter process."}. Nowadays, going to higher orders in the chiral expansion,  one can compute the process with a theoretical uncertainty less than 1\% in the amplitude (see e.g., \cite{PMR}).\footnote{In the modern parlance of chiral perturbation theory, if one computes the amplitude to,  say,  the next-to-next-to-leading order ($N^2LO$)  in the thermal np capture process, one will get the same result automatically. There will be no need to invoke low-energy theorems etc. The only thing that one will miss in such an approach is an intuitive understanding of what's going on that is supplied by the low-energy theorems. It is this intuitive insight that led to the prediction that the isovector axial  exchange charge transition is enhanced in the same way as the magnetic diople transition by the soft pions while the isovector vector charge and the Gamow-Teller axial current transitions are not~\cite{KDR,MR91}. All these are confirmed by the modern ChPT.}

Soon after the Riska-Brown work, I was invited to Stony Brook for a year in 1973, the first of many one-year visits that I have made. This was the beginning of our joint work on how chiral symmetry figures in nuclear physics\footnote{An early application of Weinberg's counting rule~\cite{weinberg}, which is now widely applied,  was discussed in \cite{ericemr} in an abridged form borrowed from Weinberg's celebrated 1979 paper on pion dynamics.}, in a sense presaging the current impressive progress made in chiral nuclear dynamics (as for instance, presented by other speakers in this meeting). The development that ensued since then up to 2008 in which Gerry played a crucial role, either directly or indirectly, is recounted in the 2011 Gerry Brown Festschrift~\cite{subtle}.

What I would like to present in this talk is a new development which has to do with the role that vector mesons play in nuclear chiral dynamics, both nuclear forces and electroweak response functions. In the approach Weinberg advocated~\cite{weinberg} which has been adopted in the developments in nuclear physics with an impressive success that ensued up to date, one writes the chiral Lagrangian with the baryon and (pseudo-Goldstone) pion fields and expands it in the systematic power expansion -- which is well-defined -- to high enough order feasible to determine all the parameters involved. In this approach, other (heavy) meson degrees of freedom such as the vector mesons are ignored or more precisely integrated out. For processes whose energy scale is much less than $m_\rho$, the mass of the lowest-lying vector meson, the systematic calculation to the relevant chiral orders should account reliably for {\em all} components of nuclear forces applicable in nuclear processes. Gerry and I had no disagreements with Weinberg on this point in dilute matter such as finite nuclei and possibly nuclear matter up to the saturation density $n_0\simeq 0.16$ fm$^{-3}$. It is possible to justify this procedure by that the mass of the lowest-vector meson ($\rho$) in free space,  770 MeV, is much greater than the pion mass and that nuclear processes probed involve energy scales much less than the vector-meson mass. It should work then in dilute nuclear matter systems probed at low energy.

The question that arises when one wants to go beyond in density is: Is there a situation in nuclear physics where the vector-meson degrees of freedom cannot be ignored? In answering this  question affirmatively, Gerry and I departed from Weinberg's view. For this we will specifically consider the role of $\rho$ meson in the nuclear tensor forces.  The pion exchange gives of course the well-defined and known pion tensor, so the question is when and where the $\rho$ meson contributes in the tensor forces in such a way that its explicit role cannot be ignored.

We approach this problem by resorting to the strategy of introducing the vector mesons in hidden local symmetry (HLS). The important point is that the Lagrangian with the vector mesons $(\rho,\omega)$ appearing as hidden local fields is gauge-equivalent to the non-linear sigma model~\cite{HLS}. Furthermore by treating in particular the $\rho$ as light as the pion, it is possible to set up a chiral expansion to recover the standard chiral perturbation theory~\cite{HY:PR}. Thus one should recover, with the vectors present explicitly, Weinberg's results -- without vectors -- in dilute nuclear systems. In this case, the HLS is nothing more than a ``philosophical" luxury. However the power of the HLS scheme that the non-linear sigma model does not possess is that one can access the regime where the $\rho$ degree of freedom cannot be integrated out. This would be the case if the $\rho$ mass were tuned to vanish together with the pion mass in the chiral limit. In free space as well as in dilute nuclear matter, $m_\rho\gg m_\pi$. However by going to high density, the vector meson mass could be made to drop sufficiently low so that the HLS becomes more applicable --treated at low order, e.g., mean field -- while the nonlinear sigma model loses its reliability or even breaks down completely. This is essentially the suggestion made in \cite{br91}. The generic scaling behavior so predicted will be dubbed ``BR scaling."\footnote{This terminology is a bit misleading because there is in practice no one universal scaling that applies to all physical observables. This point will become clear in what follows.}

Now in the spirit of the standard chiral perturbation theory (sChPT for short)\footnote{For clarity, I will refer to as ``standard" the approaches with the chiral Lagrangian with pion field only.}, one could think of extending HLS Lagrangian to dense baryonic matter using the same procedure employed in sChPT for normal nuclear matter. As long as there are no phase transitions in the density range considered, this extrapolation procedure  should be applicable. In fact, it should work as well as the sChPT does up to $n_0$. Beyond $n_0$, however, HLS should be more predictive as the theory can encode then the unique property of HLS matched to QCD, i.e., the vector manifestation (VM), which states~\cite{HY:PR} that in the chiral limit as the quark condensate $\la\bar{q}q\ra$ is dialled to zero,
\be
m_\rho\sim g\sim \la\bar{q}q\ra\rightarrow 0
\ee
where $g$ is the hidden gauge coupling. This implies that if density is increased to the point $n_c$ at which chiral symmetry is restored, the $\rho$ mass should go to zero in the way specified by matching to QCD. This feature -- which does not seem to be present in sChPT -- is yet to be confirmed or falsified by experiments.\footnote{There have been some beautiful experiments performed to single out this effect in dilepton productions from heavy ion collisions which purport to zero in on the property of the $\rho$ meson in medium. The analysis made within the framework of HLS arrived at the conclusion that the effect of the VM, present within a small temperature window, is swamped by mundane (nuclear) strong interactions taking place near on-shell (``flash point") so that it would be extremely difficult -- if not impossible -- to single it out~\cite{needle}.}

Up to date there is no systematic bona-fide chiral perturbative calculation done using the HLS Lagrangian. The difficulty is that there are simply too many unknown terms in the HLS Lagrangian to control to the chiral order one wants or is required to go. One known way to exploit the power of HLS is to implement baryonic and scalar meson degrees of freedom explcitly into the HLS Lagrangian and do the mean field approximation with the parameters of the Lagrangian suitably scaling with density. This approach was discussed in \cite{song}.  One can justify this approach for baryonic systems that are in the Fermi-liquid state. This is because the mean field theory of relativistic Lagrangians with sufficient relevant degrees of freedom can describe the Landau-Migdal Fermi liquid phase of baryonic matter~\cite{matsui,friman-rho}.  The BR scaling of \cite{br91} is anchored on this premise.

I will argue below that in the presence of topology change, the mean field approach could break down.

In order to give the precise meaning to the scaling relation in the mean field approach, it is important to recognize that the scalar degree of freedom needed is a dilaton field $\chi$ that arises from spontaneous breaking of scale invariance which is in turn triggered by an explicit breaking of scale symmetry in QCD, namely the trace anomaly. The dilaton field can be implemented in the HLS Lagrangian as a ``conformal compensator field" (conformalon for short). It turns out that it is the dilaton condensate $\la\chi\ra$ going to zero that signals the vanishing of the hadron mass, not the chiral condensate vanishing as has been generally thought. Note that $\la\chi\ra=0$ implies $\la\bar{q}q\ra=0$, but the converse is not necessarily true. We will see in fact that the putative restoration does not imply all hadron masses go to zero when $\la\bar{q}q\ra$ vanishes.\footnote{In \cite{br91}, it was incorrectly assumed that $\la\chi\ra\propto \la\bar{q}q\ra$ as $\la\bar{q}q\ra\rightarrow 0$. }

Although not rigorously proven, it seems reasonable to accept that as long as the baryonic matter can be described as a Fermi liquid, the mean field approximation should be applicable. The problem then is: How can one determine whether a matter at some high density is or is not in a Fermi-liquid state?

The main thrust of this paper is that at some density above $n_0$, there can take place a phase transition. The question I need to address is whether this phase change destroys the Fermi liquid structure.
\section{Topology in dense matter: Skyrmions and half-skyrmions}\label{topology}
One approach known to access dense matter in QCD is to put, following the seminal work of Klebanov~\cite{Klebanov}, the skyrmions as baryons that arise as solitons from HLS Lagrangian on a crystal lattice. This approach is justified for large $N_c$ for which the crystal structure is shown to have the lowest energy for baryonic matter and become more reliable at higher density. In reality, $N_c$ is 3, far from infinity and the baryonic matter at normal density is more like a liquid than a crystal. So it is natural to question the reliability of the crystal structure for nuclear matter as well as dense matter. Obviously one cannot take at their face values {\em all} of the properties given by the crystal structure. One feature that we will take seriously -- and is adopted in what follows -- is certain generic topological properties of the model. Recent developments in condensed matter physics where skyrmions play an intricate and fascinating role (see, e.g., \cite{multifacet}) hints at such an approach in nuclear physics.

There may be several phase changes in the skyrmion crystal that we do not know of as the matter density is increased. At least one phase change has been well established and that is the topological change from skyrmions to half-skyrmions at some density (that I will denote $n_{1/2}$) . It has been shown more or less independently of the crystal symmetry that the most energetically favored state of skyrmions at some density is the half-skyrmion phase in which the single soliton carries half a baryon charge~\cite{halfskyrmion}. That there must be a topological phase change involving half-skyrmions in crystal as density is changed has been argued and explored extensively~\cite{parketal}. The argument was originally developed using the Skyrme model with pion fields only\footnote{Also in in a simplified model that  incorporates vector degrees of freedom but it is formulated incompletely so that it could not be trusted.}. However, the existence of such a phase change is surprisingly robust. It is observed with HLS fields~\cite{maetal} as well as in holographic dual QCD models~\cite{dyonicsalt} where an infinite tower of vector mesons are involved. It is significant to note that the transition is not characterized by any order parameters involving local fields. It is the topology change that signals the change of states.

One cannot say, without fully quantizing the system -- which at the moment is not possible -- whether such half-skyrmions are physical, propagating degrees of freedom. A pair of half-skyrmions could be strongly bound-- or confined -- to form a single skyrmion so that it seems to make no sense to give separate identities to them. Such a half-skyrmion structure appears to be already present in $\alpha$ particles (with four nucleons in eight half-skyrmions)~\cite{manton-sutcliffe}, which suggests that it could be present as a non-propagating degree of freedom already in normal nuclear matter. There are so far no indications in nuclear systems for deviations from the standard description of nuclear properties. Its effect could, however, manifest in matters denser than normal nuclear matter. Nuclear phenomenology~\cite{dong} indicates the preferred onset density is in the vicinity of $\sim 2n_0$.

Certain features of the half-skyrmion phase are found to be model-independent. For instance, they do not depend on the details of the degrees of freedom taken into account, such as vector and scalar fields. The differences can only be of quantitative nature. Among the notable happenings that exhibit the general features, to cite a few, are the following:
\begin{itemize}
\item The quark condensate in medium $\la\bar{q}q\ra^*$ averages to  zero in unit cell at $n_{1/2}$ but the condensate is locally non-zero, so there could be chiral density wave.
\item While the pions propagating in the background of skyrmion matter decay with the decay constant $f_\pi^*$ that decreases  (roughly linearly) with a decreasing quark condensate $(\la\bar{q}q\ra^*)^{1/2}$ up to the transition density $n_{1/2}$, they stop decaying  at $n_{1/2}$, with $f_\pi^*/f_\pi $ stabilizing at $\gsim 0.6$.\cite{maetal}
\item A  $K^-$ propagating in skyrmion medium undergoes a precipitous drop of order 60 MeV in mass at $n_{1/2}$  over and above the usual chiral attraction implied in chiral Lagrangians. This additional topological effect is likely to influence kaon condensation in compact-star medium~\cite{PKR,LRkaon}.
\item The mass of the proton fluctuating on top of a dense skyrmion background, which smoothly decreases with the dropping quark condensate $\la\bar{q}q\ra$ at increasing density, stops dropping at $n_{1/2}$ and stays more or less constant up to near chiral restoration~\cite{maetal}. This can be understood in terms of a large $N_c$ effect, with $m_N^*\sim\sqrt{N_c}f_\pi^*$.
\item When a neutron-rich skyrmion matter is collectively quantized, it is found~\cite{LPR} -- in large $N_c$ approximations -- that the ``symmetry energy" of the system $E_{sym}$ defined as
    \be
    E(n,\alpha)=E_0(n,\alpha=0)+E_{sym}\alpha^2 + O(\alpha^2)
    \ee
    with $\alpha=(N-P)/(N+P)$ with $N(P)$ the number of neutrons (protons), is given by
    \be
    E_{sym}\approx \frac{1}{8{\cal I}_\tau}\label{cusp}
    \ee
    where ${\cal I}_\tau$ is the isospin moment of inertia.  The most surprising feature that this formula predicts is that $E_{sym}$ has a cusp at $n_{1/2}\sim 2n_0$~\cite{LPR}. The cusp is made up of the symmetry energy that decreases as $n$ moves from $n_0$ to  $n_{1/2}$, turns over at $n_{1/2}$ and then increases above $n_{1/2}$. This feature, absent in standard nuclear physics approaches (e.g., sChPT), is robust, independent of detailed degrees of freedom, other than that of pion, taken into account.

\end{itemize}

Understanding the cusp structure of $E_{sym}$ constitutes the main content of what follows. It will be seen to lead to a nontrivial modification of the BR scaling of \cite{br91}, signaling the possibly important role of topology in nuclear physics at higher density. I will not touch on other matters listed above but focus on what it could do to the tensor forces, the structure of which led Gerry (and me) to disagree with Weinberg in the treatment of vector mesons in ChPT as I mentioned above and to the original formulation of the BR scaling with the vector mesons figuring explicitly.
\section{Topology change and parameters of HLS Lagrangian}
Equation (\ref{cusp}) is subleading in $N_c$, i.e.,  $\sim O(N_c^{-1})$ as compared with the dominant term $E_0$ of $O(N_c)$. For instance, the $O(1)$ Casimir term is missing. In order to understand the cusp, one may have to go beyond the large $N_c$ approximation.
So how does one proceed?

There are no (known) direct ways to calculate $1/N_c$ corrections to (\ref{cusp}). So we need tricks to treat the topology change. One finds in the literature several ways of accessing topology change. One is to trade in topology for boundary conditions. This is what was done in the QCD chiral bag~\cite{RGB}. Here the topological charge is traded with the quark number divided by $N_c$. In fact the topology can be completely traded away for the quark bag -- and vice versa -- via an anomaly produced by chiral bag boundary conditions. This is elegantly demonstrated by the ``Cheshire Cat Principle"~\cite{cheshirecat} which states that the quark-bag description of baryons and the solitonic description are equivalent. This strategy \`a la boundary condition is successfully exploited in condensed matter physics, notably for topological insulators~\cite{balachandran}. The other way is to ``translate" the topological effects into the parameters of the Hamiltonian~\cite{wilczek}. Here I will choose this second path in an exploratory way.

Returning to the symmetry energy, I start with the observation that the symmetry energy in standard nuclear physics is dominated by the tensor forces. Given the tensor forces, $E_{sym}$ in nuclear many theory can be fairly well given, in the vicinity of the equilibrium density $n_0$, by the simple closure formula~\cite{closure}
\be
E_{sym}\approx 12 \la V_T^2\ra/\bar{E}\label{closure}
\ee
where $\bar{E}\approx 200$ MeV is the average excitation energy mediated by the tensor forces. \begin{figure}[ht!]
\begin{center}
\epsfig{file=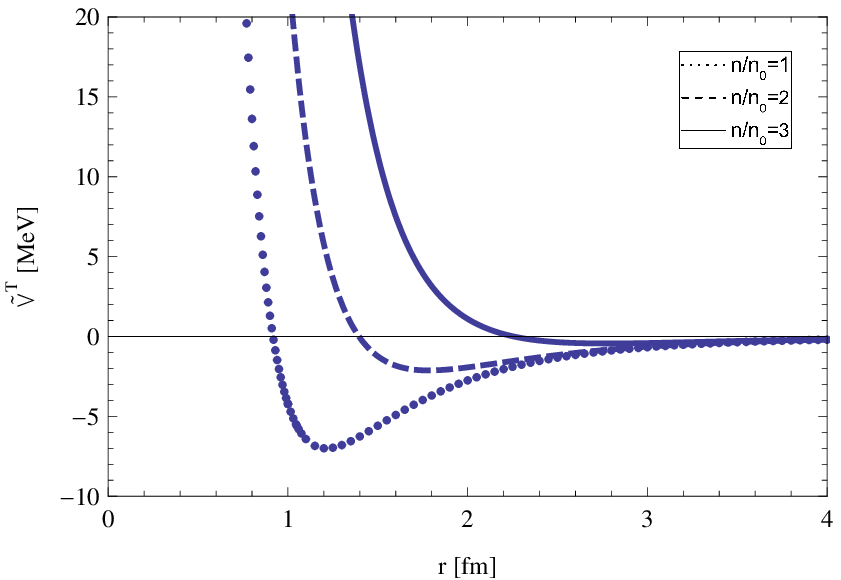,height=5.3cm}
\epsfig{file=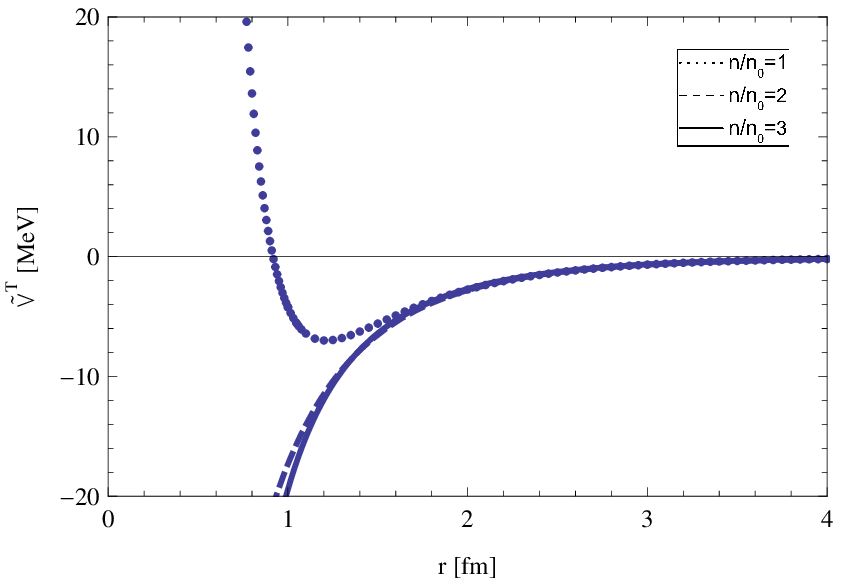,height=5.3cm}
\vskip 0cm
\caption{Schematic form of the sum of $\pi$ and $\rho$ tensor forces
(in units of MeV) for densities $n/n_0$ =1 (dotted), 2 (dashed) and 3 (solid) with the ``old scaling" (no topology change) $m_\rho^*/m_\rho \approx  1-0.15 n/n_0$ and $R\approx 1$ for all $n$ (left panel) and with the ``new scaling" (with topology change) $m_\rho^*/m_\rho \approx 1-0.15 n/n_0$  with  $R\approx 1$ for $n<n_{1/2}$ and $\Phi\equiv m_\rho^*/m_\rho\approx 1-0.15 n/n_0$ and $R\approx \Phi^2$ for $n>n_{1/2}$, assuming $ n_0<n_{1/2}<2n_0$ (right panel).}\label{tensor}
\end{center}
\end{figure}

I will now show that the formula (\ref{closure}) allows us to see how the symmetry energy with and without the topology change could behave as density increases.

In the presence of pions and vector mesons, the tensor forces are given by the sum of one-pion exchange and one-$\rho$ exchange. In HLS theory with nucleon fields explicitly implemented (called BHLS),  they take the form
\begin{eqnarray}
V_M^T(r)&&= S_M\frac{f_{NM}^2}{4\pi}m_M \tau_1 \cdot \tau_2 S_{12}\nonumber\\
&& \left(
 \left[ \frac{1}{(m_M r)^3} + \frac{1}{(m_M r)^2}
+ \frac{1}{3 m_Mr} \right] e^{-m_M r}\right),
\label{tenforce}
\end{eqnarray}
where $M=\pi, \rho$, $S_{\rho(\pi)}=+1(-1)$. It is important to note that the tensor forces come with an opposite sign between the pion and $\rho$ tensors.

It is found to be a reasonable approximation to have the pion tensor unaffected by the density for the range of density we are concerned with. I shall therefore leave it unchanged\footnote{Given the present poor knowledge of the scaling behavior of various parameters of the Lagrangian, this is the best we can do for the moment. But one should take into account the possibility of the pion tensor varying as a function of density for more realistic treatments. At the present stage of ChPT, this should be doable.}.  As for the $\rho$ tensor, apart from the scaling mass $m_\rho^*$,  there is the scaling of $f_{N\rho}$ that needs to be taken into account. Written in terms of the parameters of the baryon HLS Lagrangian (BHLS), the scaling figures in the ratio
\be
R\equiv \frac{f_{N\rho}^*}{f_{N\rho}}\approx \frac{g_{\rho NN}^*}{g_{\rho NN}}\frac{m_\rho^*}{m_\rho}\frac{m_N}{m_N^*}\ .
\ee

First consider the case of no topology change. In this case, the scaling relations given in \cite{br91} apply for all density.
\be
R&\approx& 1.\label{R}
\ee
This means that the scaling manifests entirely in the radial function of the $\rho$ tensor forces through the $\rho$ mass.\footnote{As clarified in \cite{BR:DD}, the scaling in \cite{br91} was put forward before the vector manifestation (VM) was discovered~\cite{HY:PR}.   Taking into account the VM that the $\rho$ mass scales to zero near the fixed point as the  chiral condensate does would make the relation (\ref{R}) invalid in approaching the VM fixed point. I am ignoring in Fig.~\ref{tensor} this correction which I consider not significant for the discussion.} For a  simple parametrization described in the figure caption, the net tensor forces behave as shown schematically in the left panel of Fig.\ref{tensor}: As density increases, the net tensor force strength effective in nuclear processes -- with the short-range correlations cutting off short-distance interactions -- gets weaker and disappears nearly completely at $n\gsim 2n_0$. Thus in the approximation (\ref{closure}), the symmetry energy will drop continuously, going to zero at some density. This will give what is known in the literature as ``supersoft symmetry energy."
Such scenario is produced in certain phenomenological models with a particular choice of parameters without affecting the properties at normal density, and although unorthodox it cannot be ruled out yet by compact-star phenomenology.  What is clear, however, is that the cusp given by the skyrmion-half-skyrmion transition cannot be captured by a BHLS Lagrangian without changing the parameters of the Lagrangian.

It should be stressed that the decreasing tensor forces at increasing density does get a compelling phenomenological support. It explains the C-14 dating in terms of the near vanishing of the Gamow-Teller matrix element at nuclear matter density caused by the decreasing tensor strength~\cite{c14}.\footnote{It turns out that the same suppression of the GT matrix element can also be obtained by (irreducible) chiral 3-body forces~\cite{3body}. But this does not invalidate the statement that the old BR scaling can explain the C-14 dating.}   However this process samples density below and near $n_0$. This clearly argues for that $n_{1/2}$ cannot be below $n_0$.

Let me now turn to the case with the topology change. One does not know where $n_{1/2}$ is located, but let me, on the basis of phenomenological  considerations~\cite{dong}, take it to be $n_{1/2}\simeq 2n_0$. It cannot be lower than or too close to $n_0$, for otherwise it will be at odds with what is empirically known around $n_0$, a notable example being the C-14 dating mentioned above. If it is too high above $n_0$, it may be irrelevant to the processes we are interested in. Furthermore it may not be trusted if it is very high since the effective theory must be breaking down at such high density.

I will continue to assume that the pionic tensor remains more or less  unaffected by density. Up to $n_{1/2}$, there is no change from the old scaling of \cite{br91}, so the ratio will remain $R\approx 1$ for $n\lsim n_{1/2}$. For $n>n_{1/2}$ in the half-skyrmion phase, however, we have from Section \ref{topology} that\footnote{The $\rho NN$ coupling $g_{\rho NN}$ differs from the $\rho\pi\pi$ (gauge) coupling by a factor $F_{\rho NN}$ defined as $g_{\rho NN}=gF_{\rho NN}$. This factor scales in density over and above the gauge coupling $g$ which is linked to the VM. I will come back to this factor in connection with the possible breakdown of Fermi liquid due to the topology change.}
\be
\frac{g_{\rho NN}^*}{g_{\rho NN}} \approx \frac{g^*}{g}, \ \ \frac{m_\rho^*}{m_\rho}\approx \frac{g^* f_\pi^*}{g f_\pi}, \ \ \frac{m_N}{m_N^*}\approx \frac{f_\pi^*}{f_\pi}
\ee
from which follows that
\be
R\approx (g^*/g)^2\approx (m_\rho^*/m_\rho)^2.
\ee
It is not known how the $\rho$ mass scales in the half-skyrmion phase. However one expects that it will be most likely controlled by the scaling of the gauge coupling constant $g$. The vector manifestation demands that $g$ go to zero as the chiral critical point is approached. Consequently, the $\rho$ tensor force will be suppressed strongly for $n>n_{1/2}$. This feature is seen in Fig.~\ref{tensor}, right panel. With the assumed scaling given in the figure caption, the tensor force gets quenched strongly for $n\gsim 2n_0$ and gets killed nearly completely at $n\approx 3n_0$. This will then make the pion take over the tensor force  beyond the density $n_{1/2}$ with negligible influence from the $\rho$ tensor. Since the symmetry energy due to the pion tensor force increases roughly linearly in density above $n_{1/2}$, the cusp is precisely reproduced .

Having deduced the scaling properties of the parameters of HLS Lagrangian, it still remains to show that the cusp is real, not an artifact of the crystal structure valid in the large $N_c$ limit. As mentioned, the symmetry energy given by (\ref{cusp}) is the leading order in $N_c$.  To show that the cusp can be present in nature, one can take the procedure developed by Gerry Brown, Tom Kuo and their colleagues at Stony Brook~\cite{veff}, which was to take an effective Lagrangian with suitable scaling parameters, do the ``double decimation" renormalization group analysis via $V_{lowK}$ and then take into account high-order nuclear correlations. One such calculation employing ring-diagram summation with $V_{lowK}$ derived from the HLS Lagrangian so obtained  was applied to the EoS of compact stars~\cite{dong}. With a suitable scaling function that is consistent up to normal nuclear matter but is a prediction of the theory for $n\gsim n_{1/2}$, the maximum mass of the star is found to come out $\sim 2.4 M_\odot$, with radius $\sim 11km$ and the maximum central density $\sim 5.5n_0$. These results are surprisingly reasonable. One potentially important  distinctive feature of the approach with the topology change encoded is that while the symmetry energy is soft below $n_{1/2}$ -- which is consistent with nuclear matter at $n\lsim n_0$, it stiffens for $n>n_{1/2}$, just the right mechanism to accommodate the large mass stars being observed in nature. A similar stiffening of the EoS is observed in the description that involves the hadronic phase crossing over to a strong-coupling quark matter~\cite{hatsuda}. An intriguing possibility is that in the similar vein as the Cheshire Cat phenomenon via the chiral bag between the soliton and the quark bag via boundary conditions, there could be a Cheshire Cat between the half-skyrmion matter and a strongly-coupled quark matter via the change in the scaling in HLS Lagrangian.\footnote{I am tempted to conjecture that the recent observation by Kaplan that chiral bag structure could be encoded in QCD~\cite{kaplan} could be exploited to establish the chain of connections I am making here.}

As anticipated, the cusp is smoothed by correlation effects that are clearly of higher order in $1/N_c$. This was observed in \cite{dong}. However, given the uncertainty in the scaling parameters and the possibility that there can figure other degrees of freedom -- such as kaon condensation and hyperons-- in the half-skyrmion phase, these resuts should be taken with caution.
\section{Breakdown of Fermi liquid in half-skyrmion phase}
Let me present one explanation as to why the change of parameters resulting from the topological transition is required in HLS for the nuclear tensor forces at high density. This is highly exploratory and needs to be sharpened.

For the purpose of this discussion, I need to have a scalar field which behaves as a chiral scalar at density near nuclear matter and a chiral four-vector $\sigma$ near chiral restoration. This can be achieved by the dilaton field $\chi$ in HLS mentioned above and demanding that nearing chiral restoration,  the $\sigma$ emerge from a field redefinition of the $\chi$ field~\cite{paeng}.  The requirement that the Lagrangian is singularity-free in the limit that $\la\bar{q}q\ra\rightarrow 0$ leads to the decoupling of $\rho$ meson from the nucleon. Written as
\be
g_{\rho NN}=gF_{\rho NN}
\ee
it is $F_{\rho NN}$ that tends to zero even if $g\neq 0$, that is before reaching the VM point. The point at which $F_{\rho NN}=0$ was referred to in \cite{paeng} as ``dialton limit fixed point (DLFP)."

Now suppose one does a mean field approximation with this Lagrangian. This is essentially what one does in the literature for relativistic mean field (RMF) calculations with (typically nonlinear) Lagrangians that possess proper symmetries. Then the symmetry energy comes from the mean field of an isovector vector field, say, the $\rho$ field. With BHLS, one has
\be
E_{sym}/n \sim (g_{\rho NN}^*)^2/(m_\rho^*)^2\sim (F_{\rho NN}^*/f_\pi^*)^2.\label{Eoverf}
\ee
As mentioned above, in the half-skyrmion phase, $f_\pi^*$ remains an unscaling constant. In approaching the DLFP, $F_{\rho NN}^*\rightarrow 0$ and hence $E_{sym}/n\rightarrow 0$. We do not know precisely where the DLFP is located but we see that we reproduce  qualitatively the same behavior as what one gets from (\ref{closure}) with the old BR scaling (no topology change). Accepting the hypotheisis-- thus far unproven -- that RMF theory is equivalent to Fermi liquid theory~\cite{matsui}, this would mean that the Fermi liquid description breaks down.\footnote{Note that in all RMF approaches currently available in the literature, the right-hand side of (\ref{Eoverf}) is taken to be a constant. That approach  has no inkling of topology changes nor of non-Fermi liquid. Note also that the same remark applies to all phenomenological models based on density functionals such as Skyrme potential models etc. Which picture is adopted in nature is of course an open issue.} This situation is reminiscent of a similar breakdown of Fermi liquid due to topology in condensed matter physics~\cite{MnSi}. A totally open question then would be: How can one approach such phenomena as kaon condensation and appearance of hyperons  -- which could be important if present -- in compact-star matter from the possibly non-Fermi liquid half-skyrmion phase?

\section{``Pristine signal" for BR scaling}
One of Gerry's (unfulfilled) dreams was to have a ``smoking-gun signal" for BR scaling. Although there are a variety of indirect evidences for -- and no evidences against -- it,  so far none has been seen in a pristine environment uncompounded by mundane background effects. I mentioned above in a footnote that in our view, heavy-ion processes have so far failed to probe the signal. I suggest one such possibility here based on tensor forces in nuclei, precisely what motivated Gerry and me to start thinking of the scaling two and half decades ago~\cite{tensorforce}.

In an elegant analysis, Otsuka and his collaborators showed that the ``bare" tensor forces are left un-renormalized in certain channels by high-order nuclear effects, such as short-range correlations, core polarizations etc.~\cite{otsuka}.  It is found in particular in the monopole matrix elements that the intrinsic  (bare) tensor forces remain unscathed by nuclear correlations whereas other components can be massively renormalized. Surprisingly it is found by Kuo~\cite{kuo} that the tensor forces appear to be {\em scale-invariant} in that the tensor components in $V_{lowK}$ obtained by the decimation from the chiral scale down to the Fermi-momentum scale at which nuclear many-body problem is studied are unchanged from the bare tensor forces. If this were true -- there is nothing to say why it could be so, then it would strongly suggest that the bare tensor force strength could be pinned down in nuclear medium, blissfully free of mundane nuclear effects, by looking at processes sensitive to the monopole matrix element such as single-particle shell evolution~\cite{otsuka}. If there is any medium effect in the tensor forces reflecting the {\it fundamental} change in the chiral condensate due to density-induced vacuum change, it could be unambiguously pinned down. If the effect of tensor forces in the monopole matrix elements can be tied down with high accuracy in the future, BR scaling could then be ``seen" in a pristine environment even in density regimes below $n_{1/2}$. This will be a precision test in nuclei, a rarity in nuclear physics. If in addition $n_{1/2}$ is close to $n_0$, RIB-type accelerators could provide a hint to this phase change associated with the topology change. At the very least, the FAIR/Darmstadt could do it.

As a final remark, let me mention a hand-written draft of paper that I received from Gerry by fax in 2007. There he was developing the idea of unifying the concepts of BR scaling, Landau-Migdal Fermi-liquid theory and Wilsonian renormalization-group in nuclear interactions. By resuscitating what is known as ``EELL" (Ericson-Ericson-Lorentz-Lorenz) effect -- which is equivalent to Landau-Migdal $g_0^\prime$ effect -- on which he was working when I first came to Stony Brook in 1973 and which Gerry was now arguing to be the strongest component of nuclear interactions arising from hidden gauge bosons, he was attempting to formulate a unified nuclear effective field theory that links the idea of 1973 to that of today. I believe this was his last work in progress before his illness stopped him, sadly left unfinished.

\subsection*{Acknowledgments}
 This note is largely based on the series of work done in collaboration with the team headed by Hyun Kyu Lee at Hanyang University in Seoul, Korea, partially supported by the WCU project of the Korean Ministry of Educational Science and Technology (R33-2008-000-10087-0).

%%%%%%%%%%%%%%%%%%%%%%%%%%%%%%%%%%%%%%%%%%%%%%%%%%%%%%%%
%%%%%%%%%%%%%%%%%%%%%%%%%%%%%%%%%%%%%%%%%%%%%%%%%%%%%%%%%

\end{document}